\documentclass[%
 reprint,
 amsmath,amssymb,aps,secnumarabic, graphics,floatfix,nofootinbib,
 tightenlines,nobibnotes,aps,prl,12pt
]{revtex4-2}
\usepackage{mathrsfs}
\usepackage{amsfonts} 
\usepackage{amssymb} 
\usepackage{float} 
\usepackage{graphicx}
\usepackage{dcolumn}
\usepackage{bm}
\usepackage{subcaption}
\DeclareCaptionJustification{justified}{\justifying}
\usepackage[colorlinks,citecolor=blue,linkcolor=blue,urlcolor=blue,bookmarks=false,hypertexnames=true]{hyperref}

\usepackage[utf8]{inputenc}
\usepackage{ragged2e}
\usepackage[font=small,justification=justified,format=plain]{caption}
\makeatletter
\def\@seccntformat#1{\csname the#1\endcsname\quad}

\makeatother
\begin{document}

\title{Image Force Effects on Tunneling Currents  in an STM - I \\  `Point charge in the Barrier Region' - Model}
\thanks{Corresponding author. Arun V. Kulkarni Tel.:0832 2580309; fax: +91-8322557033 E-mail address: avkbits@goa.bits-pilani.ac.in}%

\author{Malati Dessai}
\email{msd002@chowgules.ac.in}
 \affiliation{%
 Department of Physics, Parvatibai Chowgule College of Arts and Science, Margao, Goa, 403602.\\}
 
\author{Arun V. Kulkarni}
\affiliation{%
 	Department of Physics, Birla Institute of Technology and Science-Pilani,K K Birla Goa campus, Goa, 403726.\\
 }%
\date{\today}
\begin{abstract}
In a Scanning Tunneling Microscope (STM), when a tunneling electron treated as a point charge enters the barrier region between the tip and the sample, it induces image charges on the conducting surfaces, which modifies the shape of the potential barrier it sees. In this paper, the effect of the modification in the barrier potential due to these induced charges on the tunneling current density and currents in an STM,is studied as a function of the tip-sample distance $d$ and the Bias Potential $eV_b$. The image potential is found to reduce the height and the effective width of the potential barrier, leading to a huge increase in the tunneling current densities. This huge increase (by several order of magnitudes) is however unreasonable, prompting a revisit of the assumption that the electron in the barrier region is a point particle. \\


\end{abstract}
\date{\today}
\maketitle

\section{Introduction}\label{I}
The Scanning Tunneling Microscope (STM) is in essence a pair of conducting electrodes with a separation of a few atomic diameters, across which a Bias Potential is externally applied. The space  between the two electrodes acts as a potential barrier, and electrons tunnel across this barrier to produce tunneling currents. For these currents to be large enough to be measurable, the barrier width (space between the electrodes) must be very small. The electrodes can be regarded as a combination of a tip (which is sharply pointed) and the sample, (which is supposedly relatively flat, with small hills and valleys representing the presence or absence respectively of atoms/molecules on the surface). 
\begin{figure}[h]
	\centering	\includegraphics[width=3.2in,height=1.4in]{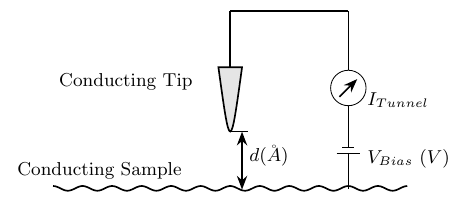}
	\caption{Schematic of a scanning tunneling microscope (STM). An external Bias Potential $V$ is applied between the metallic tip and the conducting sample.} 
	\label{fig:tip}	
\end{figure}

When the two electrodes are first brought close together, tunneling starts to occur until the Fermi levels of the two electrodes match up, and a constant Contact Potential, which is the difference in the work functions of the two electrodes, $\Delta \phi = (\phi_1 -\phi_2) $ is developed. Note the Contact Potential $\Delta \phi$ does not depend upon the Fermi energies $\eta_1$ and $\eta_2$ of the two electrodes. When a Bias Potential $eV_b$ is applied, the net potential difference between the two electrodes is $eV_b^{net} = eV_b + \Delta \phi$. For similar electrodes $\phi_1 = \phi_2$, and the the Contact Potential is zero.  The electrostatic potential in the barrier region is determined by the externally applied Bias Potential $eV_b$, the geometry of the tip specified by the radius of curvature $R$ of the tip at its lowest point, the tip-sample distance $d$, which is the distance between the nearest point of the tip and the sample, as well as the Contact Potential if the electrodes are of distinct materials. 

In the planar model of the Scanning Tunneling Microscope (STM), the surfaces of the two electrodes labeled respectively as the tip and the sample, are assumed to be two infinite parallel planes with few $\AA$  separation between them. The electrostatic potential between the tip and the sample, is linear in the distance $x$ from the emitter electrode (usually chosen to be the first electrode, or the left electrode). This function is also referred to as being trapezoidal and is given by  
\begin{equation}
 U_\text{Trap}(x) =  (\eta_1 + \phi_1) - (\phi_1-\phi_2 + eV_b) \dfrac{x}{d} 
 \end{equation}
The slope and intercepts of this function are determined by $eV_b^{net}$.  However this trapezoidal potential will be modified by image charges that are formed as soon as electrons start tunneling between the electrodes. The free electrons inside the electrodes rearrange themselves in response to the electric field created by the tunneling electrons and this rearrangement leads to induced charges appearing on the electrode surfaces, which modify the shape of the barrier potential. 

\begin{figure}[h]
	\centering
	\includegraphics[width=2.5in,height=1.7in]{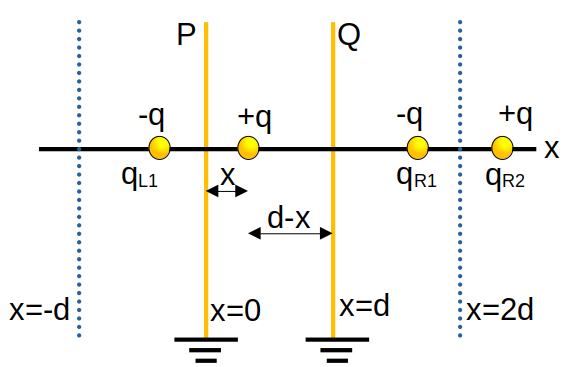}
	\caption{Image-charge construction for a point charge located between two grounded parallel conducting planes.} 
	\label{fig:img}	
\end{figure}

In this paper, the effect of the modification in the barrier potential (due to the induced charges on the conducting surfaces), on the tunneling current density in an STM, is studied, as a function of the tip-sample distance $d$ and the Bias Potential $eV_b$ in the planar model. In non-planar models (in which the tip is sharp and the sample is flat)  image effects on tunneling currents are calculated using the planar model tunneling current densities and integrating over the field lines, as was first suggested by Saenz and Garcia \cite{SG} and used in M. Dessai, and A. Kulkarni. \cite{dessai2022calculation,dessai2024thesis}.

\section{Simmons Image - Planar Model}\label{II}
\begin{figure}[h]
	\centering		
        \includegraphics[width=2.6in,height=1.9in]{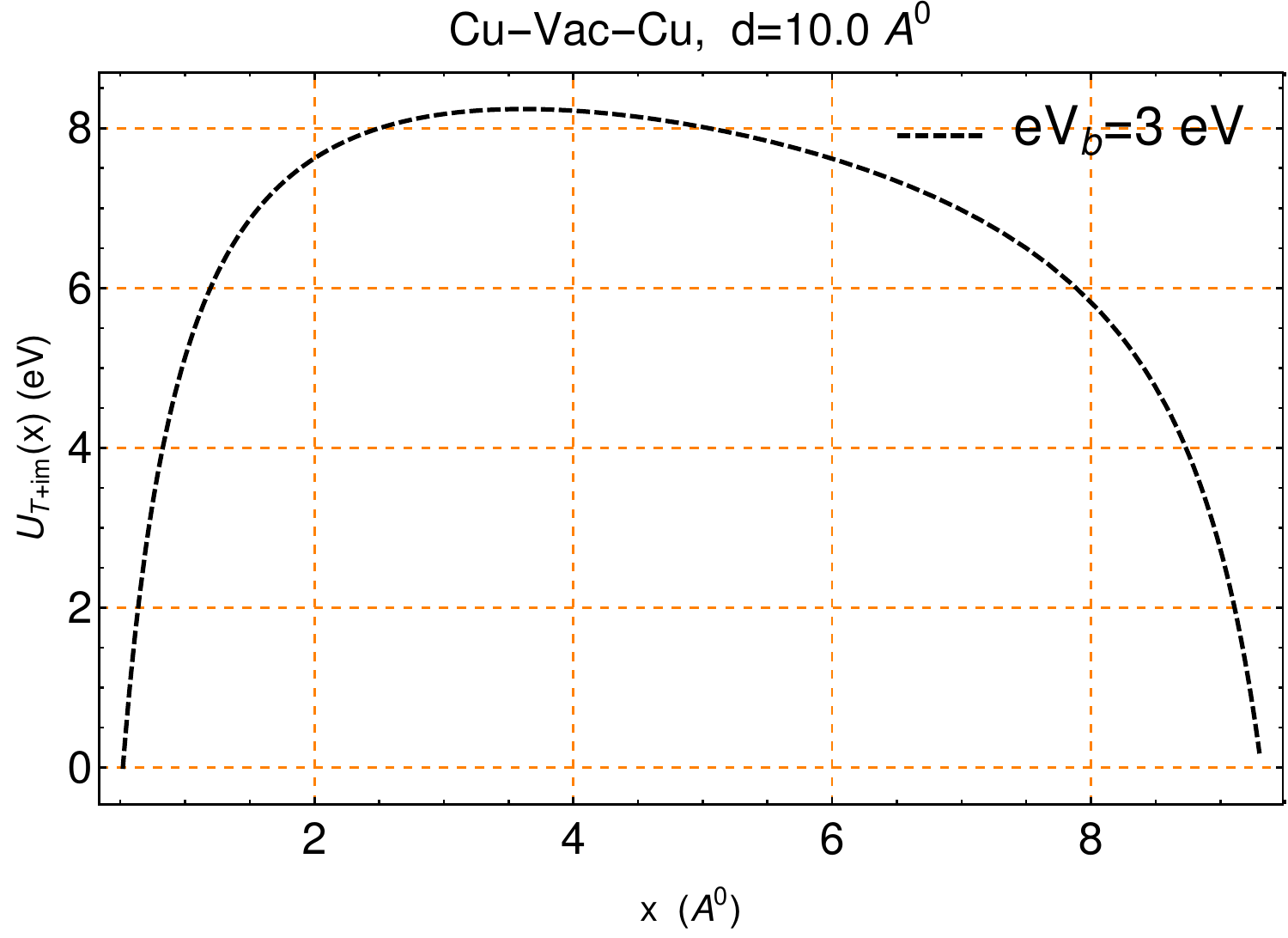}
		\caption{Plot of Trapezoid + Simmons Image Potential vs $x$ for  Bias Potential  $eV_b= 3.0$ eV.} 
		\label{fig:UTimd10V3}
\end{figure}

One of the earliest uses of the method of images to estimate the modification of the trapezoidal (linear) potential due to induced charges, was carried out by J. G. Simmons \cite{simmonsI}. He assumed that an electron behaves like a point particle lying between the two electrodes, during the tunneling process. The net potential is the sum of the trapezoidal potential and the potential due to an infinite number of images \cite{morse} formed in the interior of the two electrodes. The sum of the potentials due to these infinite image charges evaluated at the location  of the point electron (which is the source charge) inside the barrier region, is the image potential. It is given by
\begin{equation}\label{Uim_mulImg}
U_{im} = -\frac{q_e^2}{4 \pi  \epsilon_0 }\left(\sum_{n=1}^{\infty} \left[\frac{n\,d}{(n\,d)^2 - x^2}-\frac{1}{n\,d}\right] \right)
\end{equation}
where $x$ is the distance of the electron from electrode 1, and $q_e = -|e|$ is the charge of the electron and $d$ is the distance between the two electrodes. Simmons \cite{simmonsI} approximated the infinite sum due to image charges and cast the sum into a more manageable form given by the following expression called  Simmon's Image Potential (SIP).
\begin{equation}\label{Uim_approx}
U^S_{im}(x) = - 1.15 \dfrac{q_e^2 \, \ln(2)}{8\pi \epsilon_0 d} \dfrac{d^2}{x(d-x)} 
\end{equation}
 The shape of the functions in eqns (\ref{Uim_mulImg}) and (\ref{Uim_approx}) are remarkably similar as shown by Simmons \cite{simmonsI}. The tunneling electron thus \textit{sees} a net  potential (no longer trapezoidal) which is the sum of the original trapezoidal barrier potential  and the Simmons Image Potential (SIP) term $U^S_{im}(x)$. This Simmons Non- trapezoidal potential is given by 
\begin{equation}\label{UTNim}
U_{Tim} =  U_\text{Trap}(x) + U^S_{im}(x) 
\end{equation}
Fig. \ref{fig:UTimd10V3}, shows the potential barrier modified by the addition of the SIP. The net potential approaches $-\infty$ at both $x = 0$ and $ x= d$.  The net potential is spread over the three spatial regions as follows. 
 \begin{equation}\label{Unet}
U(x)=\left\{
\begin{array}{@{}lll@{}}
0 & 	\quad	x < 0\  \\
U_{Tim}(x) & 	\quad	0 \leqslant x \leqslant d \\
-\Delta_B &	\quad x > d  \\
\end{array}\right.
\end{equation} 
Note 
\begin{equation}\nonumber
\begin{array}{@{}lll@{}}
x \leqslant 0 &  \text{Region I (inside electrode 1)}\\
0 \leqslant x \leqslant d & \text {Region II (in the barrier region)}\\
x \geqslant d & \text{Region III (inside electrode 2)}\\
\end{array}
\end{equation}
where $\Delta_B=(\eta_2-\eta_1+eV_b)$. 
Let  $s_{1E}$ and $s_{2E}$  be the $x-$ coordinates of the classical turning points of the potential $U_{Tim}(x)$ at the energy $E_x$ which is the kinetic energy of the electron motion along the $x  $ axis.   The classical  turning points of the pure trapezoidal potential $U_\text{Trap}(x)$ will be at  $$s_{1E} = 0 \quad \text{and} \quad s_{2E} = d$$ This will be true for all those energies belonging to the Energy Stage I given by   $$ \text{Max}[0,(\eta_1-\eta_2-eV_b) ] \, \leqslant \, E_x \, \leqslant \, (\eta_1+\phi_2-eV_b) $$
The turning points for energies in the Energy Stage II given by  $$ (\eta_1+\phi_2-eV_b) \, \leqslant \,E_x \, \leqslant \,  E_{max} $$ are$$s_{1E} = 0\quad  \text{and} \quad s_{2E} = \frac{(\eta_1 + \phi_1 -E_x)}{\Delta\phi +eV_b}d $$ where $E_{max}$ is the maximum energy  of the tunneling electron in Energy Stage II which is $(\eta_1 + \phi_1)$. However in most cases the Fermi Dirac Function which controls the Pauli effects becomes very small, well before the energy reaches this maximum value. Thus,  there isn't much contribution to tunneling from electrons whose energies exceed $(\eta_1 + 10\, k T)$, where $T$ is the temperature of both electrodes which is fixed at $300$ K.  Hence in most calculations $E_{max}$ is usually set to  $(\eta_1 + 10\, k T)$. 

The Schr\"odinger equation for the net potential needs to be solved to find the tunneling amplitude which can later be used to calculate the tunneling probabilities and thereby the tunneling current densities. However, analytical solutions to the Schr\"odinger equation for the net potentials are difficult to find, and approximation methods for solution need to be devised. Simmons \cite{simmonsI} used the net (non-trapezoidal) potential in a WKB approximation, thus obviating the need to solve the Schr\"odinger equation exactly.  Simmons \cite{simmonsI} also makes several approximations in addition to the WKB approximation, and calculates and plots tunnel resistivities. It is difficult to fully estimate the effect of the image potentials in these calculations. 

In this paper, a WKB calculation with a direct numerical integration over the distance within the barrier is carried out to determine the tunneling probability for each value of the energy of the tunneling electron. The contribution of all energies of the tunneling electrons from 0 to $E_{max}$ are included. The calculation is carried at finite temperature of the electrodes fixed at $300$ K. 

If the potential in the barrier region were constant ($U_0$)  $viz.$ a square well potential, then (for $E_x < U_0$),  one can define $\kappa  = \sqrt{\dfrac{2 m}{\hbar^2}} \sqrt{(U_0-E_x)}$.  In the barrier region $s_{1E} \,\leqslant \,x \,\leqslant\, s_{2E}$, the wave function is of the form 
\begin{equation}
    \psi_{II}(x) = A e^{\kappa x} + B e^{-\kappa x}.
\end{equation} 
The above wavefunction needs to be joined to the wave function inside  the region of the first electrode $x \,\leqslant \, s_{1E}$ which is 

\begin{equation}
    \psi_I(x) = e^{ik_1\, x} + R  e^{-ik_1\, x} 
\end{equation}
and to the wave function inside the second electrode 
\begin{equation}
    \psi_{III}(x) = Te^{ik_2\, x}
\end{equation}
where $k_1$ and $k_2$ are the wave numbers of the electron in the first and the second electrodes respectively and $R$ and $T$ are the reflection and transmission amplitudes. Eliminating $R$ between the above equations gives  
\begin{equation}
T =
\frac{2e^{-i(k_1 s_{1E} - k_2 s_{2E})}}{
(1 + \frac{k_2}{k_1})\cosh(\kappa d')
- i\left(\frac{k_2}{\kappa} + \frac{\kappa}{k_1}\right)\sinh(\kappa d')
}
\end{equation}
where $d^\prime = s_{2E}-s_{1E}$  is the effective barrier width. In this paper the WKB  approximation for this problem is incorporated by  redefining  $\kappa$  for a non-constant, variable potential such as  $U_{Tim}(x)$, as 
\begin{equation}\label{kappa}
\kappa (E_x) = \frac{1}{d^\prime}\sqrt{\frac{2m}{\hbar^2}}
\int\limits_{s_{1E}}^{s_{2E}} dx \sqrt{U_{Tim}(x) - E_x}\,
\end{equation}
The integral for $\kappa (E_x)$,  can be calculated analytically for a pure trapezoidal potential.  For the Trapezoid + SIP  $(U_{Tim}(x))$,  $s_{1E}$ and $s_{2E}$  can be seen to be the roots of a cubic equation in $x$. The corresponding integral will have to be calculated numerically. The tunneling current densities for the WKB calculation $viz$  $J_\text{Net}^\text{WKB}$ for both the pure trapezoidal potential $U_{Trap}$ as well as for $U_{Tim}$  can be calculated directly from the general relation for the net tunneling current density in terms of the  tunneling amplitudes \cite{simmonsI,dessai2022calculation} given by 
\begin{equation}\label{JNet1}
 J_{Net}=\frac{4\pi me}{h^3}\int\limits_0^{E_{max}} dE_x |T(E_x)|^2 \mathcal{F}(E_x)
 \end{equation}
where 
\begin{widetext}
\begin{equation}\label{paulib}
\mathcal{F}(E_x)= \dfrac{k_2}{k_1}F_1(E_x) - \dfrac{k_1}{k_2}F_1(E_x+eV_b) + \frac{\Big [\dfrac{k_1}{k_2}-\dfrac{k_2}{k_1}\Big ]}{(1-e^{-\beta eV_b})}\Big [F_1(E_x+eV_b)-e^{-\beta eV_b}F_1(E_x)\Big] 
\end{equation}
\end{widetext}
and
$$F_1(E_x)=\dfrac{1}{\beta}\text{ln}[1+e^{-\beta(E_x-\eta_1)}]$$ 
The factor $\mathcal{F}(E_x)$  incorporates Pauli blocking in both electrodes.\cite{dessai2022calculation}.  Here $\beta = \frac{1}{kT} $ where $T$ is the absolute temperature of both electrodes.  
\begin{figure}[h]
    \centering
    \includegraphics[width=0.85\linewidth]{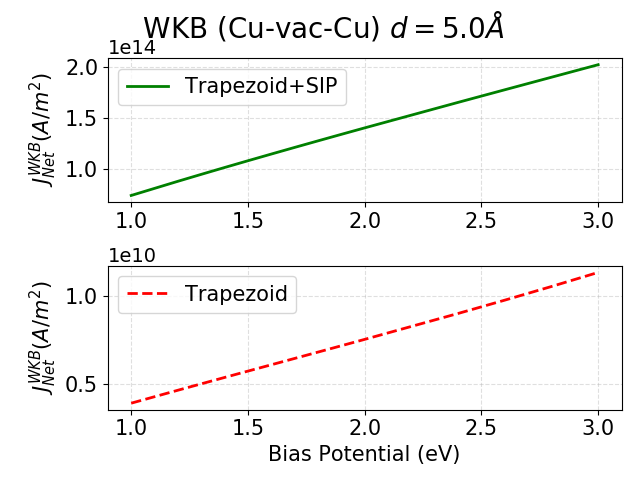}
    \caption{Plot of the WKB  tunneling current density $J_{Net}$  vs $eV_b$ for Trapezoid only and Trapezoid +SIP  for a fixed tip-sample distance $d=5 \AA.$}
    \label{fig:JwithNwithoutimg_V}
\end{figure}
 \begin{figure}[h]
     \centering
     \includegraphics[width=0.8\linewidth]{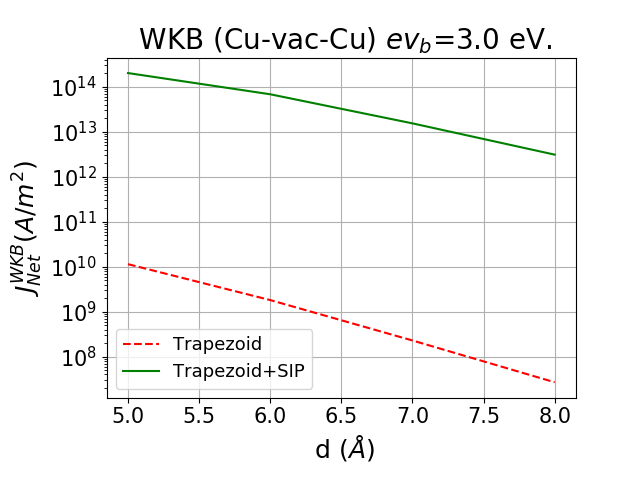}
     \caption{Plot of the WKB  tunneling current density $J_{Net}$  vs $d$ for Trapezoid only and Trapezoid + SIP  for a  fixed Bias Potential $eV_b=3.0 \, eV.$}
     \label{fig:JwithNwithoutimg_d}
 \end{figure}

In Figures \ref{fig:JwithNwithoutimg_V}  and  \ref{fig:JwithNwithoutimg_d} the WKB tunnel current densities are plotted against the applied Bias Potential $eV_b$ and the tip-sample distance $d$ respectively.  In Fig. \ref{fig:JwithNwithoutimg_V}  the tip-sample distance is fixed at $d = 5.0 \r{A}$ and in Fig. \ref{fig:JwithNwithoutimg_d} the applied Bias Potential is fixed at $eV_b=3.0 \,$eV.   

In these Figures and several Figures that follow,  the current densities J (and currents I)  in which the tunneling probabilities were calculated with the barrier potential  modified by the SIP term,  are plotted by a solid curve, while the current densities J (and currents I) with barrier potential without the SIP contribution (may be called the unmodified potential)  are plotted with a dashed line.  The Increase Factor (IF) may be defined as the ratio $$IF = \dfrac{J (I)\, \text{with the SIP Term contribution}}{J(I)\,\text{without the SIP Term contribution} }$$
The Increase Factor for current densities  in Figures \ref {fig:JwithNwithoutimg_V} and \ref{fig:JwithNwithoutimg_d}  is 4 to 5 orders of magnitude. 
 
\section{Multi - slice method for the Simmons non-trapezoid Potential (in the Planar Model)}\label{III}
To progress from the WKB approximation towards the exact solution of Schr\"odinger equation, the barrier region is divided up into very thin slices. The Simmons non-trapezoidal potential described in equation (\ref{UTNim})  is replaced by its linear approximation within each slice. The Schr\"odinger equation is now exactly solvable in each slice, the exact solutions being made up of Airy's functions $Ai$ and $Bi$. Thus the Simmons non-trapezoidal potential is modeled by a sequence of piecewise continuous linear potentials.  The corresponding exact solution in adjacent slices are joined invoking the  continuity of the wavefunction and that of its first derivative across the common boundary. The boundary points of the slices are also called the node points.

It is convenient to choose the classical turning points  $s_{10}$ and $s_{20}$  for energies  $E_x=0$, $s_{1E}$  and $s_{2E}$ for $E_x > 0$, and  $x_p$, the peak of the potential $U_{Tim}$ as node points.   Since the Simmons non-trapezoidal potential tends to $-\infty$ at $x = 0$ and  at $x = d$,  the barrier region is redefined  so that it is bounded by $s_{10}$ and $s_{20} \,\, i.e.$  $s_{10}\, \leqslant\, x \, \leqslant\,s_{20}$.  This allows one to ignore  spurious bound states near the electrode surfaces. The effective barrier width is now defined by  $d_w = s_{20}-s_{10}\, < \, d$ (the actual barrier width).   These 5 points will divide the effective barrier region into 4 sub-regions each of which is further subdivided into thin slices. The widths of these slices is chosen to be small enough, so that the potential $U_{Tim}$ is well approximated by a linear function of $x$ in each slice. 

Consider any one of the 4 sub-regions, and let it be divided by a certain number of slices of equal width by the node points $\{x_j\}, (j = 1,2, \cdots)$ so that the $j^\text{th}$ slice has endpoints at $x_j$ and $x_{j+1}(> x_j)$. Let
$$ U_j =  U_{Tim}(x_j) \, \, \text{and} \, \, U_{j+1} = U_{Tim}(x_{j+1})$$ be the values of the net potential $U_{Tim}$ at the nodes $x = x_j$ and $x_{j+1}$. Define $$\gamma_j = \dfrac{U_{j+1}-U_j}{x_{j+1}-x_j}$$ The linear interpolant of $U_{Tim}(x)$ in the $j^\text{th}$ slice is given by 
\begin{equation} 
U_j^L(x) = U_j + \gamma_j (x- x_j).
\end{equation}
The above set of linear functions satisfy 
\begin{equation} 
U_{j-1}(x_j) = U_j(x_j)
\end{equation}  
so as to maintain continuity of the potential across the common node belonging to neighboring slices. This procedure is carried out in each of the four sub-regions. Note that the width of the slices in one sub-region may not be equal to that in another sub-region.
    
The solution to the Schr\"odinger equation in the $j^{th}$ slice  $x_j \leqslant x \leqslant  x_{j+1}$ will be of the form 
\begin{equation}
\psi_j(x) = C_j Ai\big[h_j(x)\big] + D_j Bi\big[h_j(x)\big] 
\end{equation}
where $Ai \, \text{and} \, Bi$ are the Airy functions and  $C_j, \,   D_j$  are constants.  
$$h_j(x) = \frac{A_j}{B_j^{2/3}}-B_j^{1/3}x$$
\begin{equation}\nonumber
A_j=\dfrac{2m}{\hbar^2} (U_j-\gamma_j x_j-E_x),\,\, B_j=\dfrac{2m}{\hbar^2} \gamma_j
\end{equation}	

Define the following  $j^{th}$ slice-functions as 
\begin{equation}
\begin{split}
& \phi_{jk}^{(1)} = Ai[h_j(x_k)], \quad \phi_{jk}^{(2)} = Bi[h_j(x_k)] \\
& {\phi_{jk}^{(1)}}^\prime = \dfrac{d}{dx} Ai[h_j(x_k)], \quad {\phi_{jk}^{(2)}}^\prime = \dfrac{d}{dx} Bi[h_j(x_k)]
\end{split}
\end{equation}
The wave functions $\psi_{j-1}(x)$ and $\psi_{j}(x)$,  and their first derivatives, at the node $x_j$ are matched.  Rearrange the resulting equations to form a matrix relation that connects the coefficients $ C_{j}\,,\,D_{j}$ of the functions $Ai$ and $Bi$ in the $j^{th}$ slice with the corresponding coefficients $C_{j-1}\,,\, D_{j-1}$ in the $({j-1})^{th}$ slice.
\begin{equation}
    \begin{bmatrix}
        C_{j}\\\\C_{j} 
    \end{bmatrix}=
\bar{\bar{M}}_j
\begin{bmatrix}
    C_{j-1}\\\\D_{j-1} 
\end{bmatrix} 
\end{equation}
where 
\begin{equation}
\bar{\bar{M}}_j =  \begin{bmatrix}
        \phi_{j,\,j}^{(1)}& \phi_{j,\,j}^{(2)} \\ \\
        \phi_{j,\,j}^{(1)'}& \phi_{j,\,j}^{(2)'}
\end{bmatrix}^{-1}
\begin{bmatrix}
    \phi_{j-1,\,j}^{(1)}& \phi_{j-1,\,j}^{(2)} \\ \\
    \phi_{j-1,\,j}^{(1)'}& \phi_{j-1,\,j}^{(2)'}
\end{bmatrix}
\end{equation}
The $2\times2$ matrix $\bar{\bar{M}}_j$ may be called the 'transfer matrix' at the node $j$.  There will thus be one such transfer matrix for each internal node starting from $j=1$ to $j=N-1$. There will also be a transfer matrix  $\bar{\bar{M}}_0$  at the node $j=0$ which corresponds to the effective edge of the $1^{st}$ electrode at $s_{10}$ and  likewise a transfer matrix  $\bar{\bar{M}}_N$ at the node  $j=N$ which corresponds to  the effective edge of the $2^{nd}$ electrode at $s_{20}$. 

The constants $C_{N-1}\, ,\,  D_{N-1}$ in the $(N-1)^{th}$ slice   $(x_{N-1} < x < s_{20})$ are related to $C_{0}\, ,\,  D_{0}$ in the $0^{th}$ slice $(s_{10} < x <  x_1)$ by a matrix $\bar{\bar{M}}$ formed by the product of all transfer matrices taken in reverse order. Thus 

\begin{equation}\label{TrnMat}
    \begin{bmatrix}
        C_{N-1}\\\\D_{N-1} 
    \end{bmatrix} = \bar{\bar{M}}
    \begin{bmatrix}
        C_{0}\\\\D_{0} 
    \end{bmatrix} 
\end{equation}   
where 
\begin{equation}
\bar{\bar{M}}= \bar{\bar{M}}_{N-1} \cdots \bar{\bar{M}}_3 \cdot \bar{\bar{M}}_2 \cdot \bar{\bar{M}}_1
\end{equation}
The transfer matrix at $s_{10}$ connects the wavefunction $\psi_0(x)$ in the $0^{th}$ slice to the wave function $\psi_I$ inside the first electrode which is made up of plane waves, where 

\begin{equation}\label{psi12}
\psi_I(x) = e^{ik_1x} + R e^{-ik_1x}\,
\end{equation}
where $k_1$ is the wave number of the electron inside the first electrode, and $R$ is the reflection amplitude.  The conditions of continuity of the wave function and its derivative at $x = s_{10}$, lead to the following matrix equation
\begin{equation} \label{eqn13}
\begin{bmatrix}
	 e^{ik_1 s_{10}} &  e^{-ik_1 s_{10}}\\ \\
	 e^{ik_1 s_{10}} &  -e^{-ik_1 s_{10}}
     \end{bmatrix}\begin{bmatrix}
	1\\\\R 
\end{bmatrix} =
     \begin{bmatrix}
     \phi_{0,\,0}^{(1)}  &  \phi_{0,\,0}^{(2)}  \\ \\
	 \frac{\phi_{0,\,0}^{(1)^\prime}}{ik_1} &  \frac{\phi_{0,\,0}^{(2)^\prime}}{{ik_1}}
\end{bmatrix}
\begin{bmatrix}
	C_0\\\\D_0 
\end{bmatrix} 
\end{equation}
 which implies that
 \begin{equation}\label{TMats10}
   \begin{bmatrix}
	C_0\\\\D_0
\end{bmatrix} = \bar{\bar{M}}_0 
      \begin{bmatrix}
	1\\\\R 
\end{bmatrix} 
\end{equation}
where
\begin{equation}
    \bar{\bar{M}}_0 =  \begin{bmatrix}
     \phi_{0,\,0}^{(1)}  &  \phi_{0,\,0}^{(2)}  \\ \\
	 \frac{\phi_{0,\,0}^{(1)^\prime}}{ik_1} &  \frac{\phi_{0,\,0}^{(2)^\prime}}{{ik_1}}
\end{bmatrix}^{-1} \begin{bmatrix}
	 e^{ik_1 s_{10}} &  e^{-ik_1 s_{10}}\\ \\
	 e^{ik_1 s_{10}} &  -e^{-ik_1 s_{10}}
     \end{bmatrix}
\end{equation}
is the transfer matrix at the $0^{th}$ node $viz$ at $x = s_{10}$. Note the primes represent derivatives with respect to $x$. Eliminating R from the matrix equation (\ref{TMats10}) above, 
\begin{equation}\label{addAts10}
2e^{ik_1 s_{10}} = C_0 M_C + D_0 M_D
\end{equation}
where 
\begin{equation}\label{MCMD}
 M_C =  \phi_{0,\,0}^{(1)} +\dfrac{\phi_{0,\,0}^{(1)^\prime}}{ik_1}\, \,, \quad  
 M_D =  \phi_{0,\,0}^{(2)} +\dfrac{\phi_{0,\,0}^{(2)^\prime}}{ik_1}   
\end{equation}
The transfer matrix at $s_{20}$ connects the wavefunction $\psi_{N-1}(x)$ in the $(N-1)^\text{th}$ slice to the wave function $\psi_F$ inside the second electrode, where 
$$  \psi_F(x) = T e^{ik_2 x}$$
The conditions of continuity of the wave function and its first derivative at $x = s_{20}$ lead to the following equations for the transmission amplitude $T$
$$ Te^{ik_2s_{20}} = C_{N-1}\phi_{N-1,\,N}^{(1)}+ D_{N-1}\phi_{N-1,\,N}^{(2)}$$
$$ Te^{ik_2s_{20}} = \frac{C_{N-1}}{ik_2}\phi_{N-1,\,N}^{(1)^\prime}+ \frac{D_{N-1}}{ik_2}\phi_{N-1,\,N}^{(2)^\prime} $$
\noindent The above equations can be written in matrix form as 
\begin{equation}\label{Teq}
T
\begin{bmatrix}
	1\\\\1 
\end{bmatrix} 
=e^{-ik_2s_{20}} \bar{\bar{M_N}}\begin{bmatrix}
	C_{N-1}\\\\D_{N-1} 
\end{bmatrix}
\end{equation}
where 
\begin{equation}
 \bar{\bar{M_N}}=\begin{bmatrix}
	\phi_{N-1,\,N}^{(1)} & \phi_{N-1,\,N}^{(2)}\\ \\
	\dfrac{1}{ik_2}\phi_{N-1,\,N}^{(1)^\prime} & \dfrac{1}{ik_2}\phi_{N-1,\,N}^{(2)^\prime}
\end{bmatrix}
\end{equation}
is the transfer matrix at the  $N^{th}$ node $viz$ at  $x = s_{20}$.  Equations (\ref{TrnMat}) and  (\ref{Teq}) give
\begin{equation}
T
\begin{bmatrix}
	1\\\\1 
\end{bmatrix} = e^{-ik_2s_{20}} 
\bar{\bar{V}}\begin{bmatrix}
	C_{0}\\\\D_{0} 
\end{bmatrix}
\end{equation}
where 
\begin{equation}\label{VMatrix}
    \bar{\bar{V}} = \bar{\bar{M_N}} \cdot \bar{\bar{M}}
\end{equation}
In terms of $V_{ij}\, (i\,,\,j=1,2)$, the matrix elements of $\bar{\bar{V}}$, the tunneling amplitude $T(E_x)$  is given by
\begin{equation}\label{ats20}
\begin{split}
 T (E_x)= e^{-ik_2s_{20}}\big[ V_{11} C_0 + V_{12} D_0 \big] \\ =  e^{-ik_2s_{20}}\big[ V_{21} C_0 + V_{22} D_0 \big]
\end{split}
\end{equation}
Solving the equations (\ref{addAts10}) and (\ref{ats20}) for the coefficients $C_0$ and $D_0$  gives  
\begin{equation}
\left.
\begin{split}
C_0 =& \dfrac{2 e^{ik_1 s_{10}} (V_{22} - V_{12})}{M_D(V_{11} - V_{21}) + M_C(V_{22} - V_{12})} \\[10pt]
D_0 =& \dfrac{2e^{i k_1 s_{10}} - M_C C_0}{M_D}
\end{split}
\right\}
\end{equation}

Substitute the values of $C_0$ and $D_0$ from the above equation into the first of equations (\ref{ats20}), to obtain $T(E_x)$ for the barrier potential given by $U_{Tim}\, \, viz.$ Trapezoid + SIP. The net current density $J_{Net}$ is obtained as in equation (\ref{JNet1}). This method of constructing  transfer matrices has been widely used in literature.  \cite{Tang2025_DoubleBarrierTunneling, Dehghani2025_UnifiedTMM, Pereyra2021_TransferMatrixMethodTheory, Umesaki2023_ProtonTransferTMM,Figueroa2025_DeltaBarrierTMM,Sebbar2023_SuperlatticeTMM, Wang2014_GrapheneBarrierTMM}. 

\begin{figure}[hpt]
	 \centering		
        \includegraphics[width=3.0in,height=2.3in]{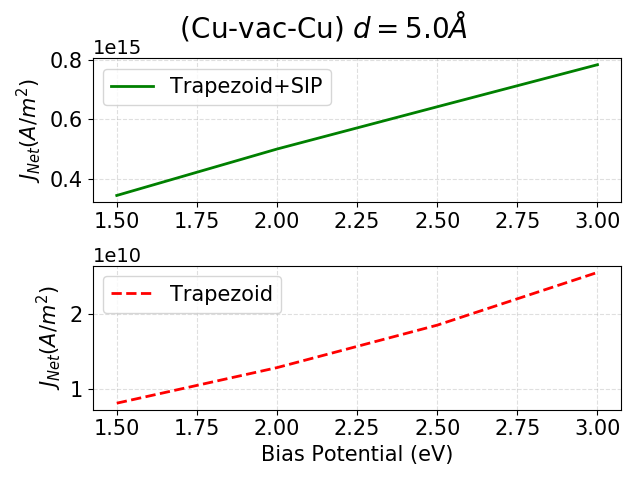}
 	\caption{Plot of current density $J_{Net}$ vs Bias Potential for Trapezoid only,  and Trapezoid+SIP for the tip-sample distance $d = 5 $ \r{A} }
	 \label{fig:JnetTimd10}
\end{figure}

\begin{figure}[hpt]
  		\centering
 		\includegraphics[width=3.0in,height=2.3in]{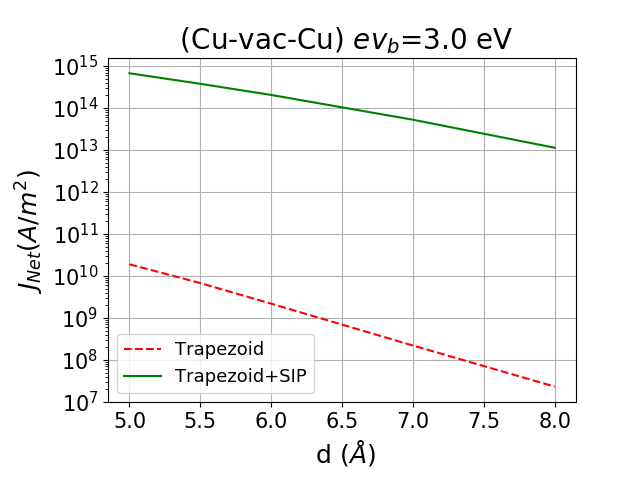}	 		
 	\caption{Plot of $J_{Net}$ vs  $d$ for Trapezoid only, and Trapezoid + SIP for  Bias Potential $eV_b= 3.0 \,$eV. }
 	\label{fig:JnetTimV4}
 \end{figure}

The current densities as calculated above are plotted as functions of the Bias Potential $eV_b$  (Fig. \ref{fig:JnetTimd10}) and the tip-sample distance $d$ (Fig. \ref{fig:JnetTimV4}).
As before solid (dashed) curve shows the $J_{Net}$ for barrier with (without) SIP. 
 \section{The Russell Potential and its modification by the Simmons Image Potential}\label{IV}
The previous section described how tunneling amplitudes are calculated in the planar model which are then used to calculate the tunneling current densities.  However,  tunneling current ( which is the real measurable quantity of interest in an STM) cannot be  calculated in planar models. To obtain the  tunneling currents in an STM, one must consider realistic tip and sample shapes. A full 3-dimensional calculation with curved surfaces, would be quite complicated and computationally expensive.  Future work in this direction is on the anvil. 

Planar models are easy to calculate and provide  essential insight into the tunneling process. Therefore these  current densities are retained,  and the method of  Saenz- Garcia \cite{saenz1994near} is used to calculate tunneling currents for actual tip and sample shapes, without requiring a full 3 dimensional calculation, using only planar model current densities. The essence of  the Saenz- Garcia \cite{saenz1994near}  method is to treat the realistic STM as being made up of infinitesimal Planar Model STM's laid along the field lines. The planar model current densities in these infinitesimal STM's is integrated over their infinitesimal cross sectional areas, to find the total tunneling current for the non-planar STM.

This task is made much easier if the tip and the sample shapes are coordinate surfaces of the same orthogonal curvilinear coordinate system. This would ensure the surfaces of both the tip, and  the sample,  would belong to the same family of surfaces, and the the field lines would lie in coordinate surfaces orthogonal to these. If the Laplace equation is separable in this coordinate system, then rotational symmetry about the $z$-axis would allow it to be reduced to a one dimensional equation in the coordinate that specifies the equipotential surfaces.  

A very convenient (for this purpose) coordinate system would be the prolate spheroidal coordinate system \cite{morse,russel,moon,cutler}  specified by 
\begin{equation}
x=\rho \cos\phi, \quad y=\rho \sin\phi, \quad  z=a\xi \eta
\end{equation}
where $\rho=a \sqrt{\xi^2-1} \sqrt{1-\eta^2}$,  and $a$ is a constant.
The ranges of the coordinate values are given by
\begin{equation}\label{range}
1\leqslant \xi \leqslant \infty, \quad -1\leqslant \eta \leqslant 1, \quad  0\leqslant \phi \leqslant 2\pi
\end{equation}
  In this system, both the tip and the sample surface would be confocal hyperboloids and the field lines would lie on the prolate spheroid. The tip surface is a surface of revolution, about the $z-$ axis of a hyperbola, and the sample surface  is along the $z=0$ plane.  Each prolate spheroid is characterized by the value of the coordinate $\xi$.  These are surfaces of constant $\xi$, and they are orthogonal to the surfaces of constant $\eta$.  If the surfaces of constant $\eta$ are chosen to be equipotential surfaces, then the lines of constant $\xi$ and $\phi$ represent electrostatic field lines. 

Due to the rotational symmetry in the system about the $z- $ axis, the potential must be independent of the azimuthal angle $\phi$. If boundary conditions are specified on the confocal hyperboloids, which are equi-potentials, then the electrostatic  potential $\Phi$ is only a function of $\eta$, and is independent of $\xi$ and $\phi$. The Laplace Equation for $\Phi$ becomes one dimensional in $\eta$, and is given by \cite{moon}
$$\frac{\partial }{\partial \eta }\left[(1-\eta ^2)\frac{\partial \Phi }{\partial \eta }\right]=0 ; \quad \eta \in [\eta_s,\, \eta_{tip}]$$
whose solution is found analytically to be 
\begin{equation}\label{solPhi}
\Phi(\eta)=\frac{A}{2}  \ln \left(\frac{1+\eta}{1-\eta }\right)+B
\end{equation}
where $A$ and $B$ are constants, to be determined by the boundary conditions imposed on $\Phi$ at the sample surface $\eta = \eta_{s}$ ($\eta_s$ is usually zero for a flat surface) and the tip surface  $\eta = \eta_{tip}( < 1)$.  The potential energy of an electron in the electrostatic potential $\Phi(\eta)$ given by equation (\ref{solPhi}) is $U(\eta) = q_e \Phi(\eta)$.  The Boundary conditions on $U(\eta)$ are 
\begin{equation}\label{33a}
\begin{split}
& U(\eta = \eta_{tip}) = \eta_1 + \phi_1\quad \text{and} \\ & U(\eta = 0) = \eta_1+\phi_2 - eV_b
\end{split}
\end{equation} 
where the $\eta_s = 0$ is describes a flat sample surface. 

Using the boundary conditions on $U$ specified in equation (\ref{33a}), the potential energy becomes 
\begin{equation}\label{17}
U(\eta) = (\eta_1+\phi_2-eV_b) + (\phi_1-\phi_2+eV_b)\, \dfrac{\lambda(\eta)}{\lambda_{\eta_{tip}}}
\end{equation}
where $$\lambda (\eta) = \ln \left (\dfrac{1+\eta}{1-\eta} \right)
\,,\,$$ and
$$\lambda_{\eta_{tip}} = \lambda(\eta_{tip}) \quad \text{and} \quad \lambda_s = \lambda(\eta_s) = 0$$
The expression for $U(\eta)$ described above, is called the Exact Russell Potential. This potential function was first introduced by A. M. Russell \cite{russel} to calculate electron trajectories in a field emission microscope.

The length along the field line stretching from $(\eta ,\xi)$ to $(\eta_{tip}, \xi)$  in which the coordinate $\xi$ is fixed, is calculated to be 
\begin{equation}
x_{fl}(\eta,\xi) = \frac{d}{\eta_{tip}} \int\limits_{\eta}^{\eta_{tip}}d\eta' \sqrt{\frac{\xi^2-\eta'^2}{1-\eta'^2}}
\end{equation}
 and 
 \begin{equation}
  d_{fl}(\xi) = x_{fl}(0, \xi)    
 \end{equation}
 is the length of the entire field line extending from the tip surface to the sample surface. The Exact Russell Potential $U(\eta)$ can also be expressed as a function of the distance  $x_{fl}(\eta, \xi)$ along a field line characterized by fixed $\xi$.  The graph of the Exact Russell Potential must pass through the two points $$[(x = 0, \implies  \eta = \eta_{tip}), \,U = (\eta_1+\phi_1)],\ \text{and}$$  $$[(x = d_{fl}(\xi) \implies \eta = 0 ),\, U = (\eta_1+\phi_2-eV_b) ]$$ as dictated by the boundary conditions in equation (\ref{33a}). The equation of a  straight line joining these two points defines the linearized Russell Potential which is then given by  \begin{equation}\label{LinRussPot}
U^L(x_{fl}) = (\eta_1+\phi_1) - (\phi_1-\phi_2+eV_b)\, \dfrac{x_{fl}(\eta,\xi)}{d_{fl}(\xi)}
\end{equation}
Note that this potential is trapezoidal. However its slope is different for different field lines.  

\begin{figure}[h]
	\centering
	\includegraphics[width=2.9in,height=2.50in]{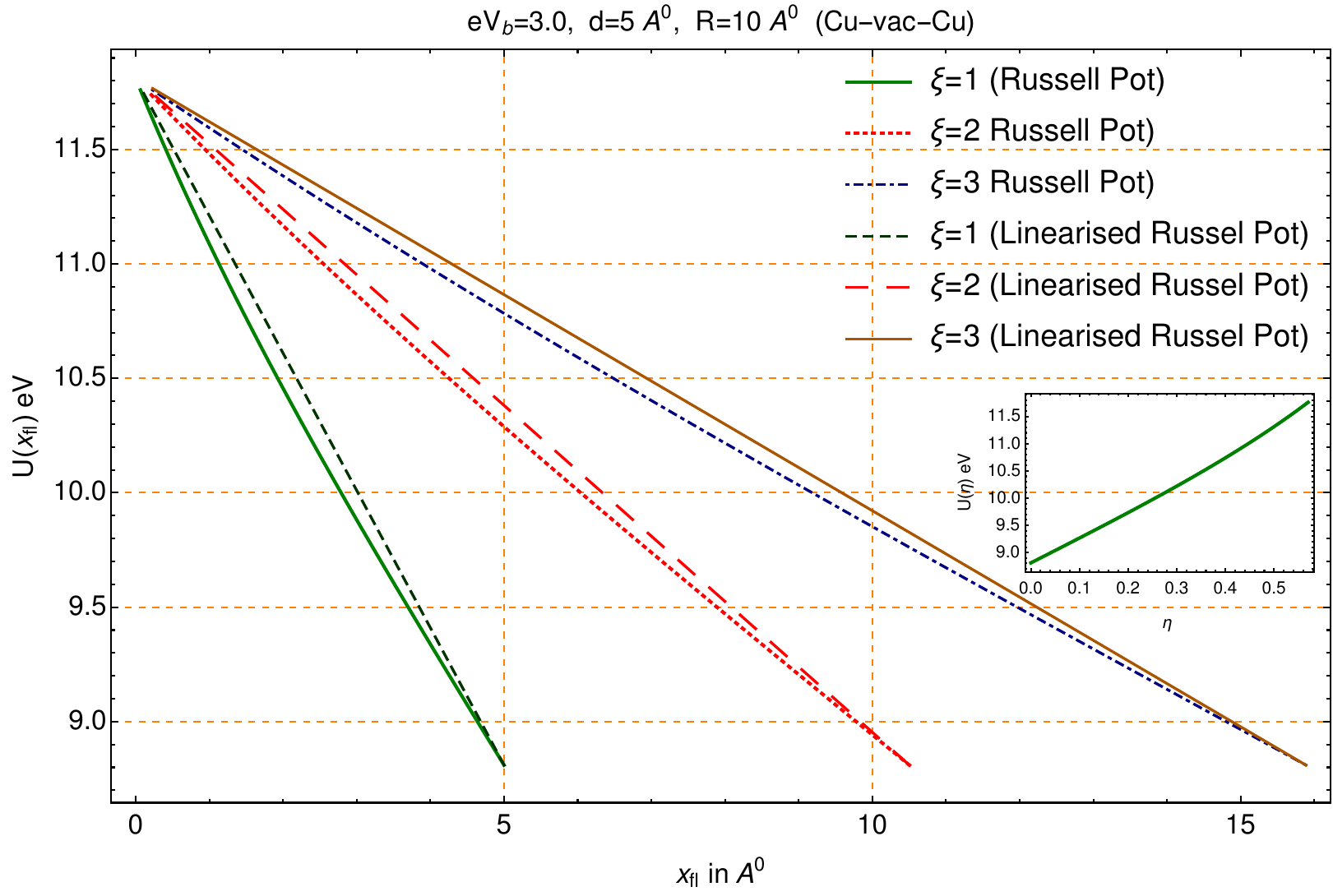}
	\caption{Plots of the Russell Potential and its linearized version as a function of $x_{fl}$ for $\xi = 1,2,3$. The inset shows the behavior of  the Russell Potential $U$ as a function of $\eta$. } 
	\label{fig:RussellNlinear}
\end{figure}

Fig. \ref{fig:RussellNlinear} shows the plots of the Exact and Linearized Russell Potential as a function of $x$,  where $x \, \epsilon\, [0,d_{fl}]$ for different values of $\xi$. The inset in this Figure shows the plot of Exact Russell potential as a function of $\eta$.  The Exact Russell Potential shows slight concave curvature, but doesn't depart very much from a straight line representing the Linearized Russell Potential.
The Exact Russell potential is therefore very nearly trapezoidal, but not exactly so.  The image corrected Exact Russell potential is given by 
\begin{equation}\label{RussExact}
U_{Net}(x_{fl}) = U(\eta) - \dfrac{1.15 q_e^2 \, \ln(2) d_{fl}}{8\pi \epsilon_0 x_{fl}(d_{fl} - x_{fl})}
\end{equation}
where $U(\eta)$ is given by equation (\ref{17}).
The inclusion of the image potential modifies the barrier potential so that each tunneling electron for every energy $E_x \,\geqslant \,0$, encounters two classical turning points.  The location of the classical turning points $s_{1E}$ and $s_{2E}$ for the energy $E_x$, for the Exact Russell Potential function $U_{Net}$ can be found by using the Brent's root finding method \cite{Brent1973}.  

\begin{figure}[hpt]
	\centering
	\includegraphics[width=3.1in,height=1.9in]{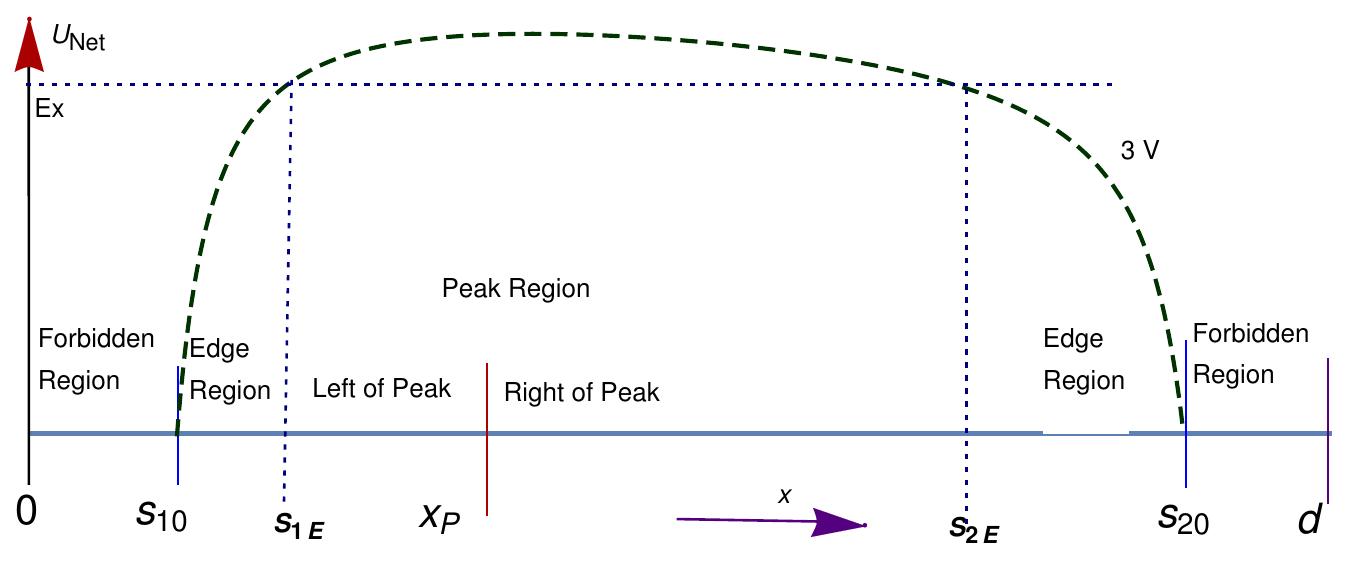}	
	\caption{Division of the Spatial Region between $[0,\,d]$ into Forbidden regions, Edge regions, and left and right of Peak regions.} 
	\label{fig:Reg}
\end{figure}

Fig. \ref{fig:Reg} shows the Exact Russell Potential as a function of the distance $x$ as measured from the tip to the sample, for a fixed Bias Potential of $3.0$ eV. It also displays the locations of the turning points $s_{10}$, $s_{20}$ and $s_{1E}$, $s_{2E}$, and the location of the peak $x_p$. It also displays the 4 sub-regions between $s_{10}$ and $s_{20}$.

 Thin slices are constructed within each sub-region, such that the $U_{Net}(x_{fl})$ which is Exact Russell + SIP described in equation  (\ref{RussExact}) can be replaced by its linear approximation. As before, the solution to the Schr\"odinger equation for a linearized $U_{Net}$ in each of the slices will be  linear combination of the Airy Functions $Ai$ and $Bi$. The coefficients of these functions in each slice can be related to those in the neighboring slices by transfer matrices as described in section \ref{III}. From these transfer matrices, tunnel amplitude $T(E_x)$ and current density $J_{Net}$ can be calculated in a manner that is completely analogous to that described in section \ref{III}. 
\section{Results and Conclusion} 
\begin{figure}[hpt]
 	 	\centering
	 	\includegraphics[width=3.0in,height=2.3in]{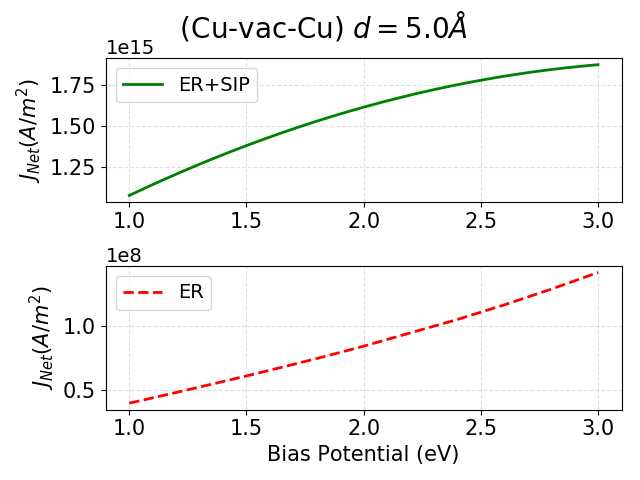}
 	\caption{Plot of $J_{Net}$ vs  $eV_b$ for Exact Russell only, and Exact Russell+ SIP for Tip-sample distance $d = 5\, \AA$ }
    \label{fig:JnetRim_d5}
	 \end{figure}  
\begin{figure}[hpt]
 	 	\centering
	 	\includegraphics[width=3.0in,height=2.3in]{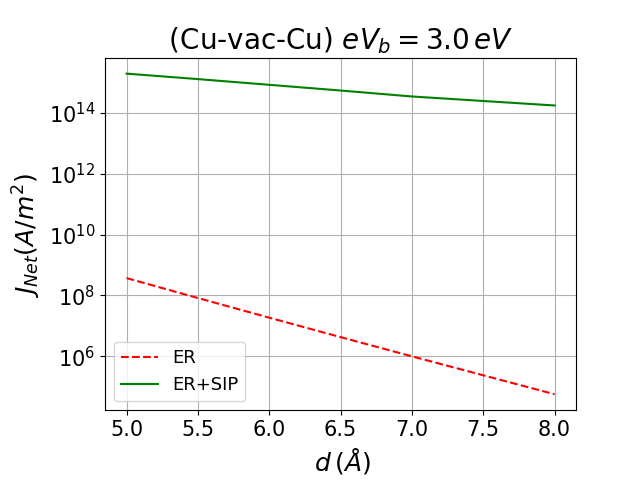}
 	\caption{Plot of $J_{Net}$ vs  $d$ comparing Exact Russell and Exact Russell+SIP for  Bias Potential $eV_b= 3.0\,$eV. }
    \label{fig:JnetRim_V3}
	 \end{figure}

Figures \ref{fig:JwithNwithoutimg_V}  to  \ref{fig:JnetTimV4}
and  Figures \ref{fig:JnetRim_d5} and \ref{fig:JnetRim_V3} display the current density as a function of the Bias Potential  and the tip-sample distance $d$ respectively.  In  these Figures, the barrier potential is trapezoidal+ SIP term (solid curve) and pure trapezoidal (dashed curve). In Figures \ref{fig:JwithNwithoutimg_V} and \ref{fig:JwithNwithoutimg_d}, the WKB approximation is used to calculate the current density for a barrier potential  with the SIP term (solid curve), and without (dashed curve). In Figures \ref{fig:JnetTimd10} and \ref{fig:JnetTimV4} the transfer matrix method is used to calculate the current density in which barrier potential is trapezoidal  and trapezoid + SIP terms (solid curve) and pure trapezoidal (dashed curve).  In Figures \ref{fig:JnetRim_d5} and \ref{fig:JnetRim_V3} the transfer matrix method is used to calculate the current density for the barrier potential with the  Exact Russell + SIP terms (solid curve)  and the Exact Russell alone (dashed curve) respectively. 

In Figures \ref{fig:JwithNwithoutimg_V} and  \ref{fig:JwithNwithoutimg_d} the SIP corrected current densities (solid curve) are larger than those without (dashed curve) by about 4 to 5 orders of magnitude. The increase factor (IF) for these Figures is therefore about 4 to 5 orders of magnitude. The same feature is found in Figures  \ref{fig:JnetTimd10} and \ref{fig:JnetTimV4} where the IF is about 4 to 5 orders of magnitude. In the Figure \ref{fig:JnetRim_d5} (\ref{fig:JnetRim_V3}), the IF is about  7 (9) orders of magnitude. The above results showing large values of the Increase Factors have been obtained with similar electrodes only.  

\begin{figure}[hpt]
	\centering
	\includegraphics[width=3.0in,height=2.50in]{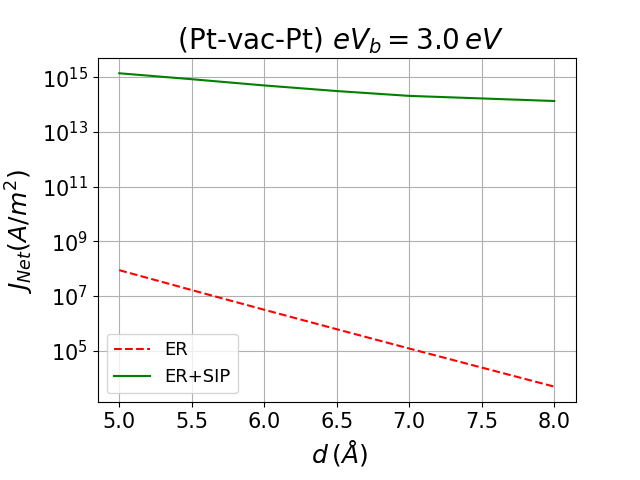}
	\caption{Plot of $J_{Net}$ vs  $d$ for Exact Russell only, and Exact Russell+SIP for Bias Potential $eV_b = 3.0$ eV. ,for similar electrodes Pt-vac-Pt.} 
	\label{fig:PtPt}
\end{figure}

\begin{figure}[hpt]
	\centering
	\includegraphics[width=3.0in,height=2.50in]{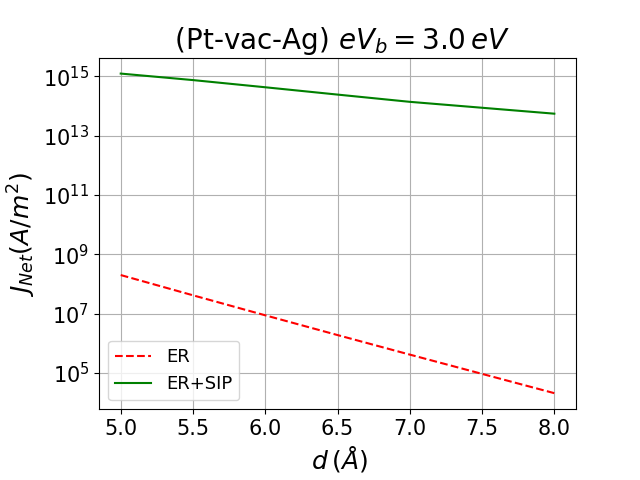}
	\caption{Plot of $J_{Net}$ vs  $d$ for Exact Russell only, and Exact Russell+SIP for Bias Potential $eV_b = 3.0$ eV.,for dissimilar electrodes Pt-vac-Ag.} 
	\label{fig:PtAg1}
\end{figure}
\begin{figure}[hpt]
	\centering
	\includegraphics[width=3.0in,height=2.50in]{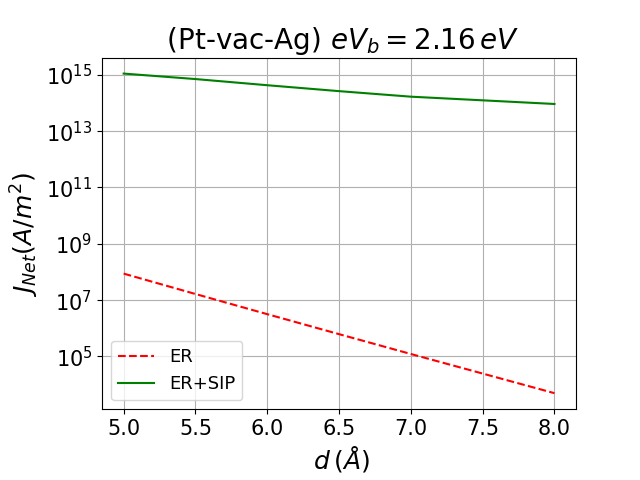}
	\caption{Plot of $J_{Net}$ vs  $d$ for Exact Russell only, and Exact Russell+SIP for Bias Potential $eV_b = (3.0$ eV. - Contact Potential), for dissimilar electrodes Pt-vac-Ag.} 
	\label{fig:PtAg2}
\end{figure}

These results  do not change much, even when dissimilar electrodes are considered.  For comparison the results ( solid and dashed curves) for the current densities for the  following electrode pairs are considered. Fig. \ref{fig:PtPt}  for similar electrodes (Platinum -vac- Platinum), with a fixed Bias Potential  of $3.0$ eV. shows the  IF   to be 7 to 10 orders of magnitude.  Fig. \ref{fig:PtAg1} plots current densities (similar legends) for dissimilar electrodes with Pt electrode as the tip and Ag electrode as the sample. In this Figure the Bias Potential is again fixed at $3.0$ eV.  The corresponding increase factor for this case is 6 to 9 orders of magnitude, which although large, is somewhat smaller than that seen in Fig. \ref{fig:PtPt}.  To make the comparison more relevant the Bias Potential in the Pt-vac-Ag system was reduced form $3.0$ eV. by the relevant Contact Potential $0.84$ eV. to $2.16$ eV., and the corresponding current densities as plotted in Fig. \ref{fig:PtAg2}. The increase factor for this Figure is 7 to 10 orders of magnitude which is the same as that for Fig. \ref{fig:PtPt}. Also the two Figures \ref{fig:PtPt} and \ref{fig:PtAg2} are remarkable similar in shape, suggesting that the results for similar electrodes are the same as for dissimilar electrodes provided the Bias Potential for the latter is corrected for the nonzero Contact Potential. Two factors are at play in this behavior. Firstly, the Contact Potential is quite small (usually less than $1.0$ eV.), and then the tunneling current density varies gradually  with the Bias Potential. Therefore the correction in the Bias Potential by the Contact Potential almost undoes the effect of dissimilarity of the electrodes.  Thus the conclusion that the increase factors are huge, holds equally well for both similar, as well as for dissimilar electrodes. 

\begin{figure}[hpt]	
	 \centering
		\includegraphics[width=3.0in,height=2.3in]{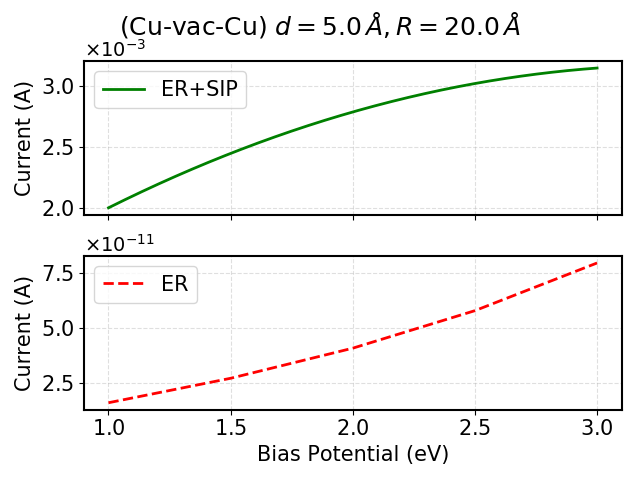}
 		\caption{Plot of current $I$ vs Bias Potential for the Exact Russell Potential only, and the Exact Russell+SIP for fixed tip-sample distance $d = 5.0 \AA$ }
	 \label{fig:I_Rimd5}
\end{figure}

\begin{figure}[hpt]	
	 \centering
		\includegraphics[width=3.0in,height=2.3in]{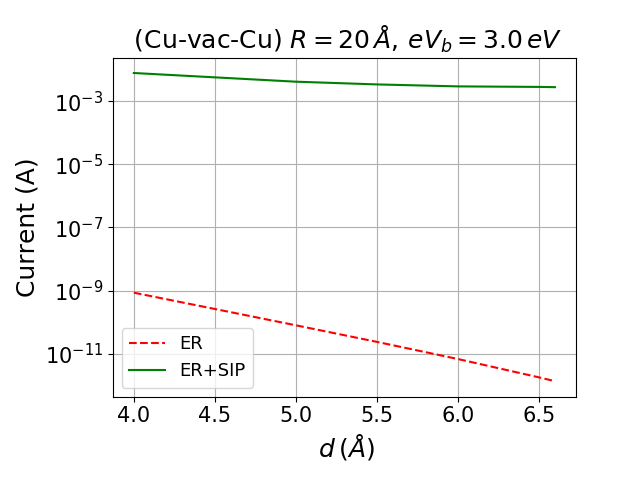}
 		\caption{Plot of current $I$ vs $d$ for the Exact Russell Potential only and the Exact Russell+SIP for fixed Bias Potential $eV_b = 3.0$ eV. }
	 \label{fig:I_RimV4}
\end{figure}

 Figures \ref{fig:I_Rimd5} and \ref{fig:I_RimV4} show tunneling currents  that are obtained from the tunneling current densities involving the Exact Russell Potential with and without the SIP term by  using the Saenz- Garcia \cite{saenz1994near} method of integrating over field lines. This method has been discussed in M. Dessai \& A. V. Kulkarni \cite{dessai2022calculation} and \cite{dessai2024thesis}. The currents so obtained show similar behavior concerning the effect of SIP term. The corresponding Increase Factors (IF) are about 8 orders of magnitude in Fig. \ref{fig:I_Rimd5} and about 7 to 9 orders of magnitude in Fig. \ref{fig:I_RimV4}. 

\begin{figure}[hpt]	
	 \centering
		\includegraphics[width=3.0in,height=2.3in]{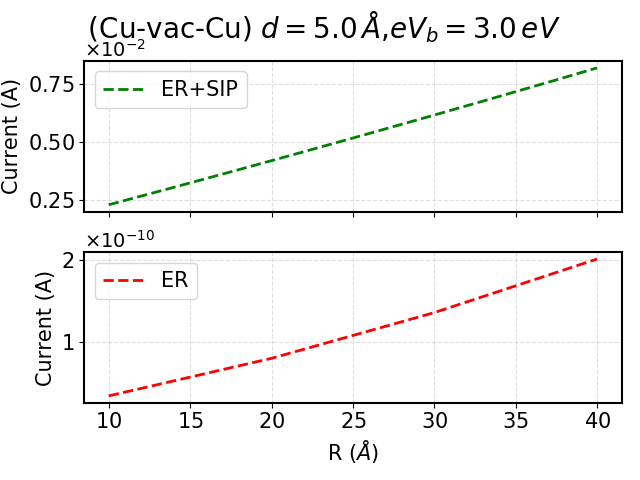}
 		\caption{Plot of current $I$ vs $R$ for the Exact Russell Potential only, and the Exact Russell+SIP for fixed tip-sample distance of $d = 5.0 \AA$ and Bias Potential  $3.0$ eV. }
	 \label{fig:I_Rimd5V4}
\end{figure}

  In Fig. \ref{fig:I_Rimd5V4}  the Bias Potential of $3.0$ eV and the  tip-sample distance of $d = 5.0 \AA$. are chosen, and tunneling currents are plotted as a function of the Radius of curvature $R$ of the tip.  The corresponding IF is about 8 orders of magnitude, for almost all values of $R$. Thus, the  non-planar aspects of the tip-sample geometry do not have much effect on the Increase Factors. Thus in all the calculations reported above, be they of current densities for Planar Models, or currents for non-planar models for several tip-radii of curvature, and where the electrodes are either similar or dissimilar, the increase factor is very large ranging from 5 to 10 orders of magnitude. This unreasonably large increase factor forces a re-examination of the principal assumptions behind these calculations.
  

The reasons for the very large values of the Increase Factors are not hard to find. Fig. \ref{fig:UTimd10V3} shows the trapezoidal potential modified by the SIP term (equation (\ref{Uim_approx})), \& Fig. \ref{fig:Reg} shows the Exact Russell potential modified by the SIP term.  In both these Figures, there are forbidden regions near the electrode surfaces, where there are spurious bound states.  To avoid dealing with these spurious bound states, the electrode  surfaces are effectively moved closer to each other. The left electrode is moved to  $x = s_{10}$ and the right electrode is moved to  $x = s_{20}$. This causes a reduction of the of the effective  barrier width.  As the energy of the tunneling electron increases, the effective barrier width gets further reduced due to  the classical turning points $s_{1E}$ and $s_{2E}$ approaching closer to each other.  Also the energy line $E = E_x$  reaches closer and closer to the peak of the barrier potential as the energy increases. Thus the tunneling electron has to climb (or tunnel through) a smaller hill at higher energies. Thus tunneling probabilities are expected to increase as the energy of the tunneling electron increases. However there is a limit to the contribution of  high energy electrons to the tunneling process. This is because the  Fermi-Dirac Factors will severely limit the  contributions of electrons whose energies greatly exceed the Fermi energy.  

Both the thinning of the barrier region,  and the reduction of the height above which the electron has to climb over (tunnel through) ,  especially at higher energies,  are the two major contributing factors leading to huge increases in the tunneling probabilities. These two factors, are absent in a pure trapezoidal or in the Exact Russell-only barrier potentials. The SIP term is therefore primarily responsible for the net barrier potential exhibiting the two features that lead to huge Increase Factors.

The huge IF's due to the SIP term, casts  doubt on  the merit of the assumption that the tunneling electron in the barrier region behaves like a point charge, which is the source of the image terms.  This assumption must be flawed  since the wave function of the tunneling electron is made up of monotonic functions only. Such a  wavefunction cannot be associated with a point particle.  Ideally the barrier region must contain a  distribution of charge whose charge density would be proportional to the square modulus of the wavefunction of the electron in this region. A correct approach would be  to solve a coupled Schrodinger + Poisson equation in the barrier region and use that solution to construct the tunneling amplitudes.  The main conclusion of this paper is therefore that, the assumption of a point electron in the barrier region causing image terms is untenable, and alternate models using distributed charges in the barrier region need to be explored, in order to correctly estimate the image contribution to the tunneling. 

\bibliography{References}

\end{document}